\documentclass[preprint,flushrt]{aastex}

\usepackage{amsmath}
\usepackage{amssymb}
\usepackage{ulem}

\slugcomment{Accepted for publication in The Astrophysical Journal} 

\shorttitle{He ENA intensities from the IBEX Ribbon}
\shortauthors{Swaczyna et al.}

\begin{document}
\title{Assessment of energetic neutral He atom intensities \\expected from the IBEX Ribbon} 
\author{P. Swaczyna, S. Grzedzielski, M. Bzowski}
\affil{Space Research Centre of the Polish Academy of Sciences, Bartycka 18A, 00-716 Warsaw, Poland}
\email{pswaczyna@cbk.waw.pl}

\begin{abstract}
 Full sky maps of energetic neutral atoms (ENA) obtained with the \textit{Interstellar Boundary Explorer} revealed a bright, arc-like Ribbon. We compare possible, though as yet undetected, He ENA emission in two models of the Ribbon origin. The models were originally developed for hydrogen ENA. In the first one, ENA are produced outside the heliopause from the ionized neutral solar wind in the direction where the local interstellar magnetic field is perpendicular to the line-of-sight. The second model proposes production at the contact layer between the Local Interstellar Cloud (LIC) and the Local Bubble (LB). The models are redesigned to helium using relevant interactions between atoms and ions. 
 Resulting intensities are compared with possible emission of helium ENA from the heliosheath. In the first model, the expected intensity is $\sim0.014~\mathrm{(cm^2~s~sr~keV)^{-1}}$, i.e., of the order of the He emission from the heliosheath, whereas in the second, the LIC/LB contact layer model, the intensity is $\sim(2-7)~\mathrm{(cm^2~s~sr~keV)^{-1}}$, i.e., a few hundred times larger. If the IBEX Ribbon needs a source population of He ENA leaving the heliosphere, it should not be visible in He ENA fluxes mainly because of the insufficient supply of the parent He ENA originating from the neutralized solar wind $\alpha$-particles. We conclude that full-sky measurements of He ENA could give promising prospect for probing the Local Interstellar Medium at the distance of a few thousand AU and create possibility of distinction between above mentioned models of Ribbon origin. 
\end{abstract}

\keywords{ISM: atoms -- ISM: magnetic fields -- ISM: bubbles -- ISM: clouds -- solar wind -- Sun: heliosphere }

\section{Introduction}
Measurements of Energetic Neutral Atoms (ENA) offer an efficient method to examine the conditions of plasma in the heliosphere, its surroundings, and in the Local Interstellar Medium (LISM). Launch of the \textit{Interstellar Boundary Explorer} (IBEX) \citep{mccomas_etal:09a} with two ENA detectors -- IBEX-Lo \citep{fuselier_etal:09a} and IBEX-Hi \citep{funsten_etal:09a} -- provided opportunity to obtain measurements of hydrogen ENA in the energy range 0.01 -- 6 keV. 

First full-sky maps \citep{mccomas_etal:09b} showed a completely unexpected feature -- a bright Ribbon, visible in the energy range 0.2 -- 6 keV \citep{funsten_etal:09b,fuselier_etal:09b}. The signal in the direction of the Ribbon is 2 -- 3 times brighter than the expected background ENA signal \citep{schwadron_etal:11a}. The Ribbon is most pronounced in the energy channels 0.7 keV, 1.1 keV, and 1.7 keV, with spatial variations possible to be explained as an 'echo' of the dynamical properties in the supersonic solar wind \citep{mccomas_etal:12b}. Recent investigation shows that the position of the Ribbon center somewhat varies with the energy channel \citep{funsten_etal:13a}. 

A few possible explanations of the IBEX Ribbon origin were presented by \citet{mccomas_etal:09b} and \citet{schwadron_etal:09a}. One of these is based on a secondary ENA production outside the heliopause. Such a model was examined in magnetohydrodynamic simulations by \citet{heerikhuisen_etal:10a,heerikhuisen_pogorelov:11a,pogorelov_etal:11a}. A simplified analytic model was recently presented by \citet{mobius_etal:13a}. Other heliospheric models of the Ribbon origin were presented by \citet{fahr_etal:11a} and \citet{fichtner_etal:13a}. 

A completely different model of the IBEX Ribbon source was proposed by \citet{grzedzielski_etal:10a}. In this model, the enhancement of the ENA intensities is explained as by production of ENA at an contact layer between the Local Interstellar Cloud (LIC) and local protrusion (bay) of the Local Bubble (LB). The ENA from the contact layer, i.e., the LIC boundary could be visible if the distance to it is smaller or comparable with the mean free path of the produced ENA. 

Helium is the second abundant species and potentially could also be prominent in the ENA fluxes. The aim of the present paper is to compare possible He ENA fluxes in the IBEX Ribbon directions in two models of the Ribbon origin: the one proposed by \citet{heerikhuisen_etal:10a} with the completely different model of extraheliospheric origin by \citet{grzedzielski_etal:10a}. As in our previous paper \citep{grzedzielski_etal:13a}, we simulated the ENA production taking into account a number of binary interactions between ions and atoms. We provide estimates of the ENA intensities in the two models of the Ribbon origin. Expected He ENA fluxes are compared with those estimated from the production in the heliosheath \citep{grzedzielski_etal:13a}.

\section{Physical models}
The models of the He ENA production we present in Section~\ref{sec:moebius} were adapted from the existing analytical models of H ENA production developed by \cite{mobius_etal:13a} and \citet{grzedzielski_etal:10a}. The first one, the secondary ENA model, explain the Ribbon as due to production located not far beyond the heliopause in the direction where the local direction of magnetic field is perpendicular to the IBEX line-of-sight. The second one (Section~\ref{sec:grzedz}), the extraheliospheric model, explains visible ENA fluxes as due to He ENA production at the hypothetic nearby contact layer between the LIC and LB, where neutral atoms from the LIC ``evaporate'' to the LB, charge exchange with ambient protons, and the products of this reaction could reach detector near the Sun.

\subsection{The secondary ENA model}
\label{sec:moebius}
In the model developed by \citet{heerikhuisen_etal:10a}, the ENA flux is produced in a sequence of three charge-exchange processes. The first one occurs in the supersonic solar wind which expands with velocity $V_\mathrm{SW}\approx440~\mathrm{km~s^{-1}}$ \citep{mobius_etal:13a}. Neutralization of the solar wind protons mainly on neutral hydrogen, inflowing from the LISM, provides a flux of neutral hydrogen of energy $\sim1~\mathrm{keV}$ escaping radially from the heliosphere, practically without further interaction, forming the so-called neutral solar wind. The mean free path of 1 keV hydrogen in the heliosheath is much larger ($\sim4\times10^{4}~\mathrm{AU}$) than the distance to the heliopause. 
Thus, reionization in the heliosheath can be neglected. Ionization of such atoms occurs only in the LISM, i.e., beyond the heliopause and is a source of Pickup Ions (PUIs) in the LISM. In the locations where the line-of-sight direction is perpendicular to the local magnetic field direction, pitch angles of PUIs are equal to 90 deg because of the radial flow of the neutral solar wind. The guiding center of such PUI is stationary. Thus neutralization of PUIs creates ENA flux in a thin disk that contains the Sun, defined by gyration of PUIs. These secondary ENA are observed as the IBEX Ribbon. This mechanism is viable if effective pitch angle scattering rate is much smaller than that of the charge-exchange. This, however, has not been ultimately confirmed \citep{florinski_etal:10a,gamayunov_etal:10a,liu_etal:12a}.

Using a formula derived by~\citet{mobius_etal:13a} it can be shown that the H ENA intensity in the IBEX Ribbon direction can be expressed as a function of mean free paths of hydrogen ($\lambda_\mathrm{H,LISM}$) and protons ($\tilde{\lambda}_\mathrm{p,LISM}$) against charge-exchange in the LISM:
\begin{equation}
 I_\mathrm{H,ENA}=\frac{\Delta\psi}{2\pi}\frac{n_\mathrm{H,SW}\left(d_\mathrm{TS}\right)V_\mathrm{SW}}{\Delta \Omega \Delta E}\frac{d_\mathrm{TS}^2}{\lambda_\mathrm{H,LISM} \tilde{\lambda}_\mathrm{p,LISM}}\underbrace{\int_{d_\mathrm{HP}}^\infty e^\frac{-\left(r_1-d_\mathrm{HP}\right)}{\lambda_\mathrm{H,LISM}} \left[ \int_{r_1}^\infty e^\frac{-\left(r_2-d_\mathrm{HP}\right)}{\lambda_\mathrm{H,LISM}} e^\frac{-\left(r_2-r_1\right)}{\tilde{\lambda}_\mathrm{p,LISM}} \frac{\mathrm{d}r_2}{r_2^2}\right] \mathrm{d}r_1}_{J\left(d_\mathrm{HP},\lambda_\mathrm{H,LISM},\tilde{\lambda}_\mathrm{p,LISM}\right)}\, ,
 \label{enaflux}
\end{equation}
where $n_\mathrm{H,SW}(d_\mathrm{TS})=8.2\times10^{-5}~\mathrm{cm^{-3}}$ \citep{mobius_etal:13a} is neutral hydrogen solar wind density at the distance of the termination shock ($d_\mathrm{TS}$), $d_\mathrm{HP}$ -- distance to the heliopause, 
$\Delta\psi=0.12$ -- FWHM of the IBEX field of view, $\Delta E\approx0.7E$ -- energy window in the appropriate energy channel of IBEX, $\Delta \Omega=0.014~\mathrm{sr}$ -- the solid angle of the IBEX field of view \citep{funsten_etal:09a,fuselier_etal:09a}. Extinction in the heliosheath was neglected, due to much longer mean free path of neutral hydrogen than the heliosheath thickness. Corrections for losses in the supersonic solar wind are performed in standard results of IBEX mission \citep{mccomas_etal:12b}. They are small for hydrogen and even smaller for He at the same energy per nucleon. The role of angles $\Delta \psi$ and $\Delta \Omega$ in this model was described by \citet{mobius_etal:13a}. The integral $J(\cdot,\cdot,\cdot)$ is dimensionless and can be calculated to amount to:
\begin{equation}
 J\left(x,y,z\right)=\frac{2z}{y-z}\exp\left({\frac{2x}{y}}\right)\mathrm{Ei}\left(- \frac{2x}{y} \right)-\frac{y+z}{y-z}\exp\left({\frac{x(y+z)}{yz}}\right)\mathrm{Ei}\left(- \frac{x(y+z)}{yz} \right)\, ,
\end{equation}
where $\mathrm{Ei}(u):=-\int_{-u}^\infty e^{-t}/t~\mathrm{d}t$. The integral $J(x,y,z)$ has a finite limit for $z\to y$.

The mean free path of neutral hydrogen solar wind against charge-exchange in the LISM depends mainly on proton density ($n_\mathrm{p,LISM}$) and charge exchange cross section at velocity $V_\mathrm{SW}$ \citep[$\sigma^\mathrm{cx}_\mathrm{H,p}=1.71\times10^{-15}~\mathrm{cm^2}$,][]{lindsay_stebbings:05a}: $\lambda_\mathrm{H,LISM}\approx (\sigma^\mathrm{cx}_\mathrm{H,p}n_\mathrm{p,LISM})^{-1}$. On the other hand, the effective mean free path, i.e., the mean path in the line-of-sight direction, of protons in the LISM is expressed as:
\begin{equation}
 \tilde{\lambda}_\mathrm{p,LISM}\approx \left( \sigma^\mathrm{cx}_\mathrm{H,p}n_\mathrm{H,LISM} \right)^{-1} \frac{V_\mathrm{LISM}\sin \theta_{BV}}{V_\mathrm{SW}}\, ,
\end{equation}
where $V_\mathrm{LISM}$ is bulk velocity of the LISM relative to the Sun, $\theta_\mathrm{BV}$ -- the angle between the direction of the interstellar magnetic field and the flow direction, $n_\mathrm{H,LISM}$ -- density of neutral hydrogen in the LISM.

\citet{mobius_etal:13a} assumed $d_\mathrm{HP}=150~\mathrm{AU}$, $d_\mathrm{TS}=100~\mathrm{AU}$, $n_\mathrm{p,LISM}=0.07~\mathrm{cm^{-3}}$, $n_\mathrm{H,LISM}=0.16~\mathrm{cm^{-3}}$, $V_\mathrm{LISM}=23.2~\mathrm{km}~\mathrm{s^{-1}}$ \citep{mobius_etal:12a}, $\theta_\mathrm{BV}=45^\circ$ \citep{pogorelov_etal:09a,opher_etal:09a} and got the intensity of H ENA $I_\mathrm{H,ENA}=470~\mathrm{(cm^2~s~sr~keV)^{-1}}$. Observed ENA intensities in the Ribbon direction at energy 0.7 (1.1)~keV are $\sim300~(200)~\mathrm{(cm^2~s~sr~keV)^{-1}}$ \citep{mccomas_etal:12b,schwadron_etal:11a}. The mean free paths for these parameters are $\lambda_\mathrm{H,LISM}\approx557~\mathrm{AU}$ and $\tilde{\lambda}_\mathrm{p,LISM}\approx9.1~\mathrm{AU}$ for energy corresponding to the solar wind velocity $V_\mathrm{SW}$.

Helium ENA production is determined by a more complex set of binary interactions. In all three steps of charge-state change, i.e., neutralization of the supersonic solar wind, ionization in the LISM and production of the secondary ENA by re-neutralization, more than one reaction is important. They are listed in Table~\ref{tab:moebius}. 

\begin{deluxetable}{cclllr}
 \tablewidth{0pt}
 \tablecolumns{6}
 \tablecaption{Reactions in the secondary ENA model \label{tab:moebius}}
 \tablehead{
    \colhead{Abbreviation $a(b)$} 		& 
    \colhead{Reaction} 				& 
    \colhead{$\sigma~[\mathrm{cm^2}]$} 		& 
    \colhead{$n~[\mathrm{cm^{-3}}]$} 		& 
    \colhead{$\sigma n~[\mathrm{cm^{-1}}]$} 	&
    \colhead{Ref.} 				

 }
 \startdata
    & \multicolumn{5}{l}{\text{1$^\circ$ Neutralization of the supersonic solar wind -- Equation~\eqref{eq:probNSW} }}\\
\noalign{\vskip 1.2ex}
\hline
2cx($\alpha$, He)	&	$\alpha + \mathrm{He} \to \mathrm{He} + \alpha$ 	&	$2.50\times10^{-16}$	&	$~0.015$	&	$3.75\times10^{-18}$	&	1	\\
cx($\alpha$, He)	&	$\alpha + \mathrm{He} \to \mathrm{He}^+ + \mathrm{He}^+$	&	$1.78\times10^{-17}$	&	$~0.015$	&	$2.68\times10^{-19}$	&	1	\\
cx($\alpha$, H)	&	$\alpha + \mathrm{H} \to \mathrm{He}^+ +\mathrm{p}$	&	$2.39\times10^{-16}$	&	$~0.08$	&	$1.91\times10^{-17}$	&	3	\\
cx(He$^+$, He)	&	$\mathrm{He}^+ + \mathrm{He} \to \mathrm{He} + \mathrm{He}^+$	&	$7.85\times10^{-16}$	&	$~0.015$	&	$1.18\times10^{-17}$	&	1	\\
cx(He$^+$, H)	&	$\mathrm{He}^+ + \mathrm{H} \to \mathrm{He} +\mathrm{p}$	&	$4.60\times10^{-17}$	&	$~0.08$	&	$3.68\times10^{-18}$	&	1	\\
\hline
\noalign{\vskip 1.2ex}
& \multicolumn{5}{l}{\text{2$^\circ$ Ionization of the neutral solar wind in the LISM -- Equation~\eqref{eq:mfpHeLISM} }}\\
\noalign{\vskip 1.2ex}
\hline			
cx(He, He$^+$)	&	$\mathrm{He} + \mathrm{He}^+ \to \mathrm{He}^+ + \mathrm{He}$	&	$7.67\times10^{-16}$	&	$~0.0096$	&	$7.38\times10^{-18}$	&	1	\\
ion(He, H)	&	$\mathrm{He} + \mathrm{H} \to \mathrm{He}^+ + \mathrm{H} + \mathrm{e}$	&	$5.27\times10^{-18}$	&	$~0.194$	&	$1.02\times10^{-18}$	&	2	\\
ion(He, He)	&	$\mathrm{He} + \mathrm{He} \to \mathrm{He}^+ + \mathrm{He} + \mathrm{e}$	&	$5.67\times10^{-18}$	&	$~0.0153$	&	$8.66\times10^{-20}$	&	1	\\
2cx(He, $\alpha$)	&	$\mathrm{He} + \alpha \to \alpha + \mathrm{He}$	&	$2.50\times10^{-16}$	&	$~0.00011$	&	$2.72\times10^{-20}$	&	1	\\
cx(He, p)	&	$\mathrm{He} + \mathrm{p} \to \mathrm{He}^+ + \mathrm{H}$	&	$4.42\times10^{-19}$	&	$~0.056$	&	$2.48\times10^{-20}$	&	1	\\
cx(He, $\alpha$)	&	$\mathrm{He} + \alpha \to \mathrm{He}^+ + \mathrm{He}^+$	&	$1.78\times10^{-17}$	&	$~0.00011$	&	$1.95\times10^{-21}$	&	1	\\
\hline
\noalign{\vskip 1.2ex}
& \multicolumn{5}{l}{\text{3$^\circ$ Production of the secondary ENA -- Equation~\eqref{eq:mfpHepLISM} }}\\
\noalign{\vskip 1.2ex}
\hline	
cx(He$^+$, He)	&	$\mathrm{He}^+ + \mathrm{He} \to \mathrm{He} + \mathrm{He}^+$	&	$7.85\times10^{-16}$	&	$~0.0153$	&	$1.20\times10^{-17}$	&	1	\\
cx(He$^+$, H)	&	$\mathrm{He}^+ + \mathrm{H} \to \mathrm{He} +\mathrm{p}$	&	$4.60\times10^{-17}$	&	$~0.194$	&	$8.93\times10^{-18}$	&	1	\\

 \enddata
 \tablecomments{Abbreviations used in Equations: 2cx(A, B)/cx(A, B) -- double/single charge exchange between A and B, ion(A, B) -- A single ionization by B impact without change of charge-state of B. The used values of cross sections, densities of the second reagent (background), and their products are presented for the energy corresponding to the solar wind velocity $V_\mathrm{SW}=440~\mathrm{km~s^{-1}}$.}
 \tablerefs{(1) \citet{redbooks}, (2) \citet{dubois_kover:89a,sarkadi_etal:13a,montenegro_etal:13a}, (3) \citet{havener_etal:05a}}
\end{deluxetable}

We assumed that upon leaving the Sun, the solar wind consists of 95\% of protons and 5\% of $\alpha$-particles by number \citep{kasper_etal:07a}. Neutralization of $\alpha$-particles in the solar wind is possible in two ways: (1) by double charge exchange with neutral interstellar helium or (2) by two single charge exchanges with neutral interstellar hydrogen and/or helium. The probability of neutralization can be approximately expressed as:
\begin{equation}
 p_{\mathrm{\alpha\to He}}\approx \int_0^{d_{\mathrm{TS}}}\left[ \sigma^\mathrm{2cx}_{\mathrm{\alpha,He}}n_{\mathrm{He}}+\left(\sigma^\mathrm{cx}_{\mathrm{He^+,He}}n_{\mathrm{He}}+\sigma^\mathrm{cx}_{\mathrm{He^+,H}}n_{\mathrm{H}}\right)\int_0^s \left(\sigma^\mathrm{cx}_{\mathrm{\alpha,He}}n_{\mathrm{He}}+\sigma^\mathrm{cx}_{\mathrm{\alpha,H}}n_{\mathrm{H}}\right)\mathrm{d} s'\right] \mathrm{d} s\, ,
 \label{eq:probNSW}
\end{equation}
where $\sigma^a_b$ denotes cross sections for appropriate reaction in part $1^\circ$ of Table~\ref{tab:moebius}, $n_\mathrm{He}$ and $n_\mathrm{H}$ are densities of neutral helium and hydrogen in the supersonic solar wind. Integral \eqref{eq:probNSW} is evaluated up to the distance to the termination shock ($d_{\mathrm{TS}}$). 
Taking a uniform distribution of neutrals in the supersonic solar wind: $n_\mathrm{H}=0.08~\mathrm{cm^{-3}}$ \citep{bzowski_etal:08a,zank_etal:13a} and $n_\mathrm{He}=0.015~\mathrm{cm^{-3}}$ \citep{witte_etal:04a,gloeckler_etal:04a,moebius_etal:04a} one gets $p_\mathrm{\alpha\to He}=0.59\%$ for $d_\mathrm{TS}=100~\mathrm{AU}$. More neutral helium is produced in the double charge exchanges ($0.55\%$) than in two single charge exchanges ($0.04\%$). Thus, the probability of $\alpha$-particle neutralization to He atom is almost proportional to the neutral helium density and the distance to the termination shock. Changes in the neutral interstellar helium densities by ionization are much smaller than in neutral hydrogen densities \citep{rucinski_etal:03}. Thus, we neglect these changes in our calculations. The small probability of neutralization obtained justifies the approximation used in Equation \eqref{eq:probNSW}.
Such a probability for He is about 35 times smaller than the probability of proton neutralization. Taking the solar wind density at 1~AU as $n_\mathrm{SW}=5~\mathrm{cm^{-3}}$ \citep{mobius_etal:13a} we get that the neutral solar wind helium density at the termination shock is $n_\mathrm{He,SW}(d_\mathrm{TS})=1.48\times10^{-7}~\mathrm{cm^{-3}}$. 

The importance of processes in the LISM is determined by products of appropriate cross sections and densities. The densities were adopted following \citet{frisch_etal:11a} and \citet{slavin_frisch:08a}, i.e., $n_\mathrm{H,LISM}=0.194~\mathrm{cm^{-3}}$, $n_\mathrm{p,LISM}=0.056~\mathrm{cm^{-3}}$, $n_\mathrm{He,LISM}=0.0153~\mathrm{cm^{-3}}$, $n_\mathrm{He^+,LISM}=0.0096~\mathrm{cm^{-3}}$, $n_\mathrm{\alpha\mathrm{,LISM}}=0.00011~\mathrm{cm^{-3}}$. For these abundances, the ionization of He from the neutral solar wind is dominated by charge exchange with He$^+$ and ionization by H impact. Other reactions, including charge exchange with protons and ionization by neutral helium, contribute less than 2\% of the total He ionization rate (cf.~part 2$^\circ$ of Table~\ref{tab:moebius}). 
We also assume that electrons are in thermal equilibrium \citep[6300 K,][]{frisch_etal:11a}, thus they have too low energy to ionize helium by impact. The mean free path against ionization of He in the LISM is approximately
\begin{equation}
 \lambda_\mathrm{He,LISM}\approx \left( \sigma^\mathrm{cx}_\mathrm{He,He^+}n_\mathrm{He^+,LISM}+\sigma^\mathrm{ion}_\mathrm{He,H}n_\mathrm{H,LISM}+\sigma^\mathrm{ion}_\mathrm{He,He}n_\mathrm{He,LISM}\right)^{-1}\approx7900~\mathrm{AU}\, .
 \label{eq:mfpHeLISM}
\end{equation}

We assume that pitch angle scattering of He PUIs is as ineffective as for H PUIs. Neutralization of the helium PUIs created by ionization of He ENA is possible due to charge exchange with H, He, and He$^+$. Cross sections for neutralization with neutrals were previously mentioned in the case of neutralization in the supersonic solar wind. Charge exchange with He$^+$ at such energies is negligible \citep{redbooks}. In that case, the effective mean free path against neutralization of the He PUIs (cf.~part 3$^\circ$ of Table~\ref{tab:moebius}) is
\begin{equation}
 \tilde{\lambda}_\mathrm{He^+,LISM}\approx \left(\sigma^\mathrm{cx}_\mathrm{He^+,He}n_\mathrm{He,LISM}+\sigma^\mathrm{cx}_\mathrm{He^+,H}n_\mathrm{H,LISM}\right)^{-1}\frac{V_\mathrm{LISM}\sin \theta_{BV}}{V_\mathrm{SW}}\approx 120~\mathrm{AU}\, .
 \label{eq:mfpHepLISM}
\end{equation}
Additional contribution from $\alpha$ PUIs is negligible because only a small amount of such ions originates from the neutral solar wind (cf. 2cx(He, $\alpha$) in part 2$^\circ$ of Table~\ref{tab:moebius}).

The intensity of He ENA could be calculated using Equation \eqref{enaflux} after replacements $\mathrm{H\to He}$, $\mathrm{p\to He^+}$, and taking value of $\Delta E\approx 0.7E$ for energy $E\approx4~\mathrm{keV}$, i.e., full energy of He ENA in this mechanism \citep[cf.][]{allegrini_etal:08a, grzedzielski_etal:13a}. This corresponds to the highest IBEX energy channel \citep{funsten_etal:09a}. In case of helium the mean free path of neutrals is also much longer than the heliosheath thickness and the losses in the supersonic solar wind are also negligible, so we neglect losses in the heliosphere. That gives the He ENA intensity $I_\mathrm{He,ENA}\approx0.014~\mathrm{(cm^2~s~sr~keV)^{-1}}$, i.e., about $3\times10^4$ smaller than the value for H ENA. 
Even taking into account that the H ENA intensity in the highest energy channel is reduced to $\sim10~\mathrm{(cm^2~s~sr~keV)^{-1}}$ \citep{mccomas_etal:12a}, the He ENA intensity is still about a factor of $10^3$ smaller than the value for H ENA. Thus the chances to detect the IBEX Ribbon in He ENA in this hypothesis are slim.

\subsection{The extraheliospheric model}
\label{sec:grzedz}

The extraheliospheric model developed by \citet{grzedzielski_etal:10a} suggests the IBEX Ribbon is a result of ENA production due to interactions at the contact layer between the LIC and LB. In this model neutrals from the LIC ``evaporate'' into the LB and neutralize a portion of the hot suprathermal plasma in the LB. The produced neutrals suffer extinction on their path to the observer both in the LB and LIC. 

The aim of \citet{grzedzielski_etal:10a} was to tune the scenario to reproduce the angular distribution of the observed IBEX Ribbon emission on the sky. An adequate combination of spatial neutral hydrogen profile in the LB at the proper angle between the line-of-sight and the interface, and the extinction in the LIC, provides angular profiles similar to those measured by IBEX \citep{funsten_etal:13a}. \citet{grzedzielski_etal:10a} considered two geometries of the interface -- planar and a spherical cap. The plane model reproduces the H ENA fluxes with good accuracy in the IBEX Ribbon direction. Comparison of the plane models for H and He should give a good approximation of the ratio of H-to-He ENA signal in the extraheliospheric model. 

\subsubsection{Ion and atom distribution}
\label{sec:grzedz:distr}

Atom and ion densities and temperatures in the LIC and the LB are crucial for analysis of the ENA signal from the LIC/LB contact layer. The densities of neutrals were taken as in Section~\ref{sec:moebius}, i.e., $n_\mathrm{H,LIC}=0.19~\mathrm{cm^{-3}}$ and $n_\mathrm{He,LIC}=0.015~\mathrm{cm^{-3}}$. \citet{grzedzielski_etal:10a} considered density of protons as a free parameter whose product with the distance to the interface determines the extinction in the LIC. The density of $\mathrm{He^+}$ ions was chosen to satisfy the condition that the ratio of H to He abundance is always 10 for every chosen value of protons density. We used the sets of ions densities in the LIC identical as \citet{grzedzielski_etal:10a}. They are presented in Table~\ref{tab:param}. 

\begin{deluxetable}{rccccccccc}
 \tablecolumns{10}
 \tablewidth{0pt}
 \tablecaption{Parameter sets in the extraheliospheric model \label{tab:param}}
 \tablehead{
    \colhead{Set} 			&	
    \colhead{$z_0$} 			&	
    \colhead{$n_\mathrm{p,LIC}$} 	&	
    \colhead{$n_\mathrm{He^+,LIC}$}	&	
    \colhead{$\tau_\mathrm{0,H}$}	&	
    \colhead{$\tau_\mathrm{0,He}$}	&	
    \colhead{$\tau_\mathrm{0,He}$}	&	
    \colhead{$v_\mathrm{ev}$}		&	
    \colhead{$f_\mathrm{p}$}		&	
    \colhead{$f_\mathrm{\alpha}$}	\\
    \colhead{}				&	
    \colhead{$\mathrm{[AU]}$}		&	
    \colhead{$\mathrm{[cm^{-3}]}$}	&	
    \colhead{$\mathrm{[cm^{-3}]}$}	&	
    \colhead{}				&	
    \colhead{$(1~\mathrm{keV})$}	&	
    \colhead{$(1~\mathrm{keV/n})$}	&	
    \colhead{$\mathrm{[km~s^{-1}]}$}	&	
    \colhead{[\%]}			&	
    \colhead{[\%]}	
 }
 \startdata
    (i)	& $	369	$ & $	0.01	$ & $	0.005	$ & $	0.10	$ & $	0.031	$ & $	0.028	$ & $	25	$ & $	0.36	$ & $	0.36	$ \\
(ii)	& $	554	$ & $	0.01	$ & $	0.005	$ & $	0.15	$ & $	0.046	$ & $	0.042	$ & $	\phn7	$ & $	1.30	$ & $	1.30	$ \\
(iii)	& $	738	$ & $	0.01	$ & $	0.005	$ & $	0.20	$ & $	0.062	$ & $	0.056	$ & $	15	$ & $	0.65	$ & $	0.65	$ \\
(iv)	& $	258	$ & $	0.03	$ & $	0.007	$ & $	0.20	$ & $	0.029	$ & $	0.025	$ & $	15	$ & $	0.65	$ & $	0.65	$ \\

 \enddata
\end{deluxetable}

Due to the high temperature of the LB, $T_\mathrm{LB}\approx10^6~\mathrm{K}$ \citep{breitschwerdt:98}, hydrogen and helium far away from the interface were assumed to be completely ionized in the form of protons and $\alpha$-particles. We adopted proton density $n_\mathrm{p,LB}=0.005~\mathrm{cm^{-3}}$ \citep{jenkins_etal:09a,welsh_shelton:09a} and set $\alpha$-particles density as $n_\mathrm{\alpha,LB}=0.1n_\mathrm{p,LB}=5\times10^{-4}~\mathrm{cm^{-3}}$. Electron density in the LB was set to $n_\mathrm{e,LB}=n_\mathrm{p,LB}+2n_\mathrm{\alpha,LB}=0.006~\mathrm{cm^{-3}}$. 

The essential binary interactions for helium are more complex than for hydrogen. The distribution of evaporated neutrals in the LB depends on the evaporation velocity and proper ionization rates in the LB. The perpendicular component of the evaporation velocity mainly determines the inflow rate of neutral atoms to the LB and their evaporation depth. However, the parallel component slightly modifies the expected relative velocity and thus the ionization rate. As in the previous paper, we assume that the evaporation velocity is perpendicular to the interface and is the same for neutral H and He, as should be expected in hydrodynamic solution \citep[cf.][and references therein]{grzedzielski_etal:10a}.

The ionization processes relevant in the extraheliospheric model are: ionization of evaporating neutrals in the LB and ionization of newly produced ENA in the LB and the LIC for extinction. In each case the total ionization rate is a sum over the rates of all relevant ionization processes: 
\begin{equation}
\gamma_{i,\mathrm{ion}}=\sum_a \alpha_a n_a\, ,
 \label{eq:ioniz_rate}
 \end{equation}
where $a$ enumerates the ionization processes for species $i$ (H or He), $\alpha_a$ is an appropriate rate coefficient, and $n_a$ -- the density of the second reagent. The rate coefficient is the expected value of the product of the cross section ($\sigma_a$) and relative velocity of reagents ($v_{\mathrm{rel,}a}$): $\alpha_a=\langle \sigma_a v_{\mathrm{rel,}a} \rangle$. The ionization rate coefficients were calculated by integrating the cross sections from references listed in Table~\ref{tab:grzedz} with velocity assuming a Maxwellian distribution for the second reagent. They are presented in parts $1^\circ$ and $2^\circ$ of Table~\ref{tab:grzedz}. 
We assume that ionization of He ENA in the LB is provided only by the completely ionized plasma, ionization by neutrals evaporating from the LIC is neglected as a second order process. The plasma in the LIC is partly ionized and thermal energy of electrons is too small to provide ionization by impact, as in the secondary ENA model.

\begin{deluxetable}{cclllr}
 \tablewidth{0pt}
 \tablecolumns{6}
 \tablecaption{Reactions rate coefficients in the extraheliospheric model \label{tab:grzedz}}
 \tablehead{
    \colhead{Abbreviation $a(b)$}		&
    \colhead{Reaction}				&
    \colhead{$\alpha_a(1.1~\mathrm{keV})$}	&
    \colhead{$\alpha_a(1.1~\mathrm{keV/n})$}	&
    \colhead{$\alpha_{a,\mathrm{ev}}$}		&
    \colhead{Ref.}				\\
						&
						&
    \colhead{$[\mathrm{cm^{3}s^{-1}}]$}		&
    \colhead{$[\mathrm{cm^{3}s^{-1}}]$}		&
    \colhead{$[\mathrm{cm^{3}s^{-1}}]$}		&
 }
 \startdata
    & \multicolumn{5}{l}{\text{1$^\circ$ Ionization in the LB} ($T_\mathrm{LB}=10^6~\mathrm{K}$)}\\
\noalign{\vskip 1.2ex}
\hline
ion(He, e)	&	$\mathrm{He} + \mathrm{e}	\to	\mathrm{He}^+ + 2\mathrm{e}$	&	$1.80	\times	10^{-8}$	&	$1.80	\times	10^{-8}$	&	$1.80	\times	10^{-8}$	&	4	\\
2cx(He, $\alpha$)	&	$\mathrm{He} + \alpha	\to	\alpha + \mathrm{He}$	&	$7.25	\times	10^{-9}$	&	$1.13	\times	10^{-8}$	&	$2.80	\times	10^{-9}$	&	1	\\
cx(He, $\alpha$)	&	$\mathrm{He} + \alpha	\to	\mathrm{He}^+ + \mathrm{He}^+$	&	$1.06	\times	10^{-10}$	&	$9.50	\times	10^{-10}$	&	$3.70	\times	10^{-12}$	&	1	\\
cx(He, p)	&	$\mathrm{He} + \mathrm{p}	\to	\mathrm{He}^+ + \mathrm{H}$	&	$8.70	\times	10^{-13}$	&	$4.11	\times	10^{-11}$	&	$7.23	\times	10^{-14}$	&	1	\\
\hline
\noalign{\vskip 1.2ex}
& \multicolumn{5}{l}{\text{2$^\circ$ Ionization in the LIC}  ($T_\mathrm{LIC}=6300~\mathrm{K}$)}\\
\noalign{\vskip 1.2ex}
\hline
cx(He, He$^+$)	&	$\mathrm{He} + \mathrm{He}^+	\to	\mathrm{He}^+ + \mathrm{He}$	&	$2.39	\times	10^{-8}$	&	$3.53	\times	10^{-8}$	&	\nodata			&	1	\\
ion(He, H)	&	$\mathrm{He} + \mathrm{H}	\to	\mathrm{He}^+ + \mathrm{H} + \mathrm{e}$	&	$3.35	\times	10^{-11}$	&	$2.63	\times	10^{-10}$	&	\nodata			&	2	\\
ion(He, He)	&	$\mathrm{He} + \mathrm{He}	\to	\mathrm{He}^+ + \mathrm{He} + \mathrm{e}$	&	$1.96	\times	10^{-11}$	&	$2.96	\times	10^{-10}$	&	\nodata			&	1	\\
cx(He, p)	&	$\mathrm{He} + \mathrm{p}	\to	\mathrm{He}^+ + \mathrm{H}$	&	$1.33	\times	10^{-13}$	&	$2.66	\times	10^{-11}$	&	\nodata			&	1	\\
\hline
\noalign{\vskip 1.2ex}
& \multicolumn{5}{l}{\text{3$^\circ$ Production of ENA in the LB} ($T_\mathrm{LB}=10^6~\mathrm{K}$)}\\
\noalign{\vskip 1.2ex}
\hline
2cx($\alpha$, He)	&	$\alpha + \mathrm{He}	\to	\mathrm{He} + \alpha$ 	&	$7.25	\times	10^{-9}$	&	$1.13	\times	10^{-8}$	&	\nodata			&	1	\\
cx($\alpha$, He)	&	$\alpha + \mathrm{He}	\to	\mathrm{He}^+ + \mathrm{He}^+$	&	$1.07	\times	10^{-10}$	&	$9.50	\times	10^{-10}$	&	\nodata			&	1	\\
cx($\alpha$, H)	&	$\alpha + \mathrm{H}	\to	\mathrm{He}^+ + \mathrm{p}$	&	$1.38	\times	10^{-9}$	&	$1.49	\times	10^{-8}$	&	\nodata			&	3	\\
cx(He$^+$, He)	&	$\mathrm{He}^+ + \mathrm{He}	\to	\mathrm{He} + \mathrm{He}^+$	&	$2.47	\times	10^{-9}$	&	$3.56	\times	10^{-8}$	&	\nodata			&	1	\\
cx(He$^+$, H)	&	$\mathrm{He}^+ + \mathrm{H}	\to	\mathrm{He} + \mathrm{p}$	&	$6.90	\times	10^{-10}$	&	$2.35	\times	10^{-9}$	&	\nodata			&	1	\\
cx(He$^+$, $\alpha$)	&	$\mathrm{He}^+ + \alpha	\to	\alpha + \mathrm{He}^+$	&	$1.68	\times	10^{-8}$	&	$2.56	\times	10^{-8}$	&	\nodata			&	1	\\
ion(He$^+$, e)	&	$\mathrm{He}^+ + e	\to	\alpha + 2e$	&	$2.19	\times	10^{-9}$	&	$2.19	\times	10^{-9}$	&	\nodata			&	4	\\

 \enddata
 \tablerefs{(1) \citet{redbooks}, (2) \citet{dubois_kover:89a,sarkadi_etal:13a,montenegro_etal:13a}, (3) \citet{havener_etal:05a}, (4) \citet{janev_etal:87a}}
\end{deluxetable}

The density of neutrals $n_i$ ($i=\mathrm{H,~He}$) in the LB can be expressed as a function of distance to the interface $y$ and velocity of evaporation perpendicular to the interface $v_\mathrm{ev}$:
\begin{equation}
 n_{i}(y)=n_{i\mathrm{,LIC}}\exp\left(-y\frac{\gamma_{i\mathrm{,ion,ev}}}{v_\mathrm{ev}}\right)\, ,
 \label{eq:dens}
\end{equation}
where the rate $\gamma_{i\mathrm{,ion,ev}}$ \eqref{eq:ioniz_rate} is a sum over ionization processes. The rates coefficients for helium are listed in the fifth column of Table~\ref{tab:grzedz}. Helium ionization is dominated by electron impact. The rates of the other proccesses are at least 100 times smaller. For hydrogen, both electron impact and charge-exchange have comparable rates \citep{grzedzielski_etal:10a}.

\subsubsection{Suprathermal ions in the LB}
\label{sec:grzedz:suprathermal}
\citet{grzedzielski_etal:10a} introduced the density ratio $f_\mathrm{p}$ of the suprathermal to thermal protons. The model intensities of H ENA in the Ribbon direction for energy $E=1.1~\mathrm{keV}$ are comparable with the ones measured by IBEX-Hi if $f_\mathrm{p}\sim1\%$. The energy width of the suprathermal population was assumed to correspond to the IBEX-Hi energy window. Such value is reasonable, e.g., for a $\kappa$-distribution \citep{livadiotis_mccomas:13a}; for temperature $T_\mathrm{LB}$ we get $f_\mathrm{p}=1.24\%$ (for $\kappa=2.5$). We include the energy window width $\Delta E$ in the ratio $f_\mathrm{p}$.

The suprathermal to thermal ratio of $\alpha$-particles ($f_\alpha$) needed to estimate the He ENA production has to be considered separately. We adopted two hypotheses: (a) $f_\alpha=f_\mathrm{p}$ for the same absolute energy range, i.e., $E=1.1~\mathrm{keV}$; (b) $f_\alpha=f_\mathrm{p}$ for the same range of energy per nucleon, i.e., $E=1.1~\mathrm{keV/n}$ (full energy of He $4.4~\mathrm{keV}$). In hypothesis (a) $\alpha$-particles and protons are taken from the same temperature distribution, e.g., in the $\kappa$-distribution. Hypothesis (b) implies that velocity distributions of protons and $\alpha$-particles are the same. The situation in hypothesis (b) is characteristic for acceleration on shocks \citep{korreck_etal:04a}. 

\subsubsection{ENA source function in the LB}
The local intensity source function depends on densities of neutrals and suprathermal populations, and appropriate cross sections. In the case of hydrogen it can be expressed as:
\begin{equation}
 j_\mathrm{H,ENA}(y)=f_\mathrm{p}n_\mathrm{p,LB}\alpha^\mathrm{cx}_\mathrm{H,p} n_\mathrm{H}(y)\, ,
 \label{eq:sourceH}
\end{equation}
where $y$ is the distance from the interface, $\alpha^\mathrm{cx}_\mathrm{H,p}$ is the rate coefficient for charge exchange between proton and hydrogen atom, for ENA energy $1.1~\mathrm{keV}$ equal to $7.92\times10^{-8}~\mathrm{cm^3s^{-1}}$ \citep{lindsay_stebbings:05a}. For helium, the formula for the source function is much more complex because the charge state of He can undergo changes by a number of different reactions:
\begin{equation}
 j_\mathrm{He,ENA}(y)=f_\alpha n_\mathrm{\alpha,LB}\left[\alpha^\mathrm{2cx}_\mathrm{\alpha,He} n_\mathrm{He}(y)+\left(\alpha^\mathrm{cx}_\mathrm{\alpha,He} n_\mathrm{He}(y)+\alpha^\mathrm{cx}_\mathrm{\alpha,H} n_\mathrm{H}(y)\right)q_\mathrm{He^+\to He}(y)\right]\, ,
 \label{eq:sourceHe}
\end{equation}
where 
\begin{equation}
 q_\mathrm{He^+\to He}(y)=\frac{\alpha^\mathrm{cx}_\mathrm{He^+,He}n_\mathrm{He}(y)+\alpha^\mathrm{cx}_\mathrm{He^+,H}n_\mathrm{H}(y)}{\alpha^\mathrm{cx}_\mathrm{He^+,He}n_\mathrm{He}(y)+\alpha^\mathrm{cx}_\mathrm{He^+,H}n_\mathrm{H}(y)+\alpha^\mathrm{cx}_\mathrm{He^+,\alpha}n_\mathrm{\alpha,LB}+\alpha^\mathrm{ion}_\mathrm{He^+,e}n_\mathrm{e,LB}}
 \label{eq:prob}
\end{equation}
is the probability of a second neutralization after first decharging  of an $\alpha$-particle, instead of its reionization. All relevant rate coefficients $\alpha^a_b$ are presented in part $3^\circ$ of Table~\ref{tab:grzedz}. Possible reactions for neutralization are the same as used in Section~\ref{sec:moebius} for neutralization of the solar wind. 

\subsubsection{ENA intensities near the heliosphere}
Using the source functions as described in Equations \eqref{eq:sourceH} and \eqref{eq:sourceHe}, the ENA intensities could be expressed by the integral
\begin{equation}
 I_{i,\mathrm{ENA}}(\theta)=\exp\left(-z_0\sec \theta\frac{\gamma_{i\mathrm{,LIC}}}{v_\mathrm{ENA}}\right)\int_0^\infty \frac{j_{i\mathrm{,ENA}}(x\cos\theta)}{4\pi}\exp\left(-x\frac{\gamma_{i\mathrm{,LB}}}{v_\mathrm{ENA}}\right)  \mathrm{d}x\, ,
 \label{eq:intensity}
\end{equation}
where $i$ denotes species (H or He), $v_\mathrm{ENA}$ -- velocity of produced ENA, $z_0$ -- distance in the direction perpendicular to the interface, $\theta$ -- angle between the direction perpendicular to the interface and the line-of-sight, $\gamma_{i\mathrm{,LIC}}$ -- loss rate for ionization of ENA in the LIC, $\gamma_{i,\mathrm{LB}}$ -- loss rate for ionization of ENA in the LB.

The loss rates in the LIC depend on assumed ion densities (cf.~Section~\ref{sec:grzedz:distr}). The loss rates were calculated from Equation \eqref{eq:ioniz_rate} and the rate coefficients are presented in parts $1^\circ$ and $2^\circ$ of Table~\ref{tab:param}. The calculated total extinction thicknesses in the LIC in the direction perpendicular to the interface ($\tau_{0,i}=z_0\gamma_{i,\mathrm{LIC}}/v_\mathrm{ENA}$) and distances to the interface are presented in Table~\ref{tab:param}. Extinction of H (He) ENA in the LIC is determined by charge-exchange with protons (He$^+$), whereas in the LB it is dominated by electron impact ionization.

\begin{figure}[ht]
  \plottwo{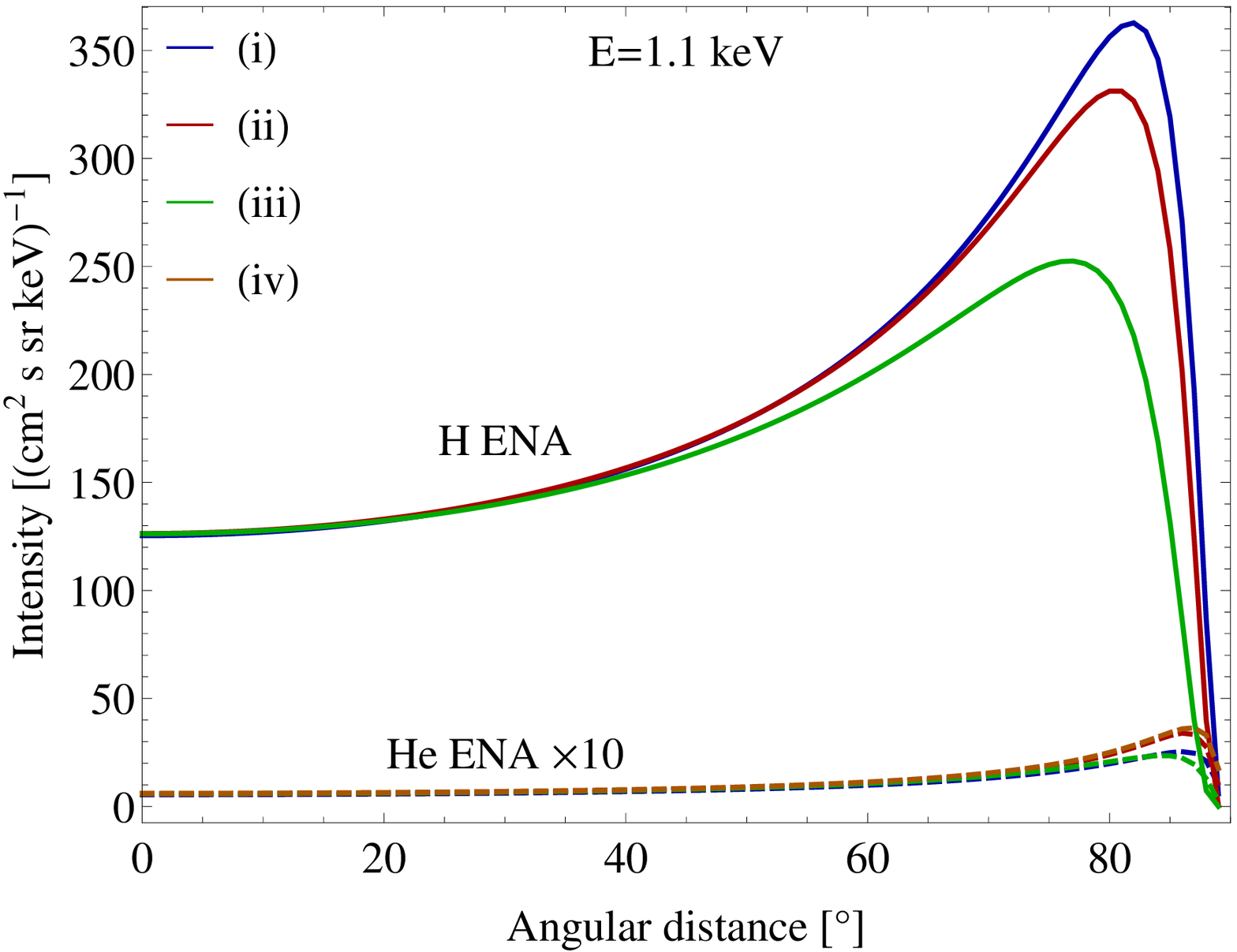}{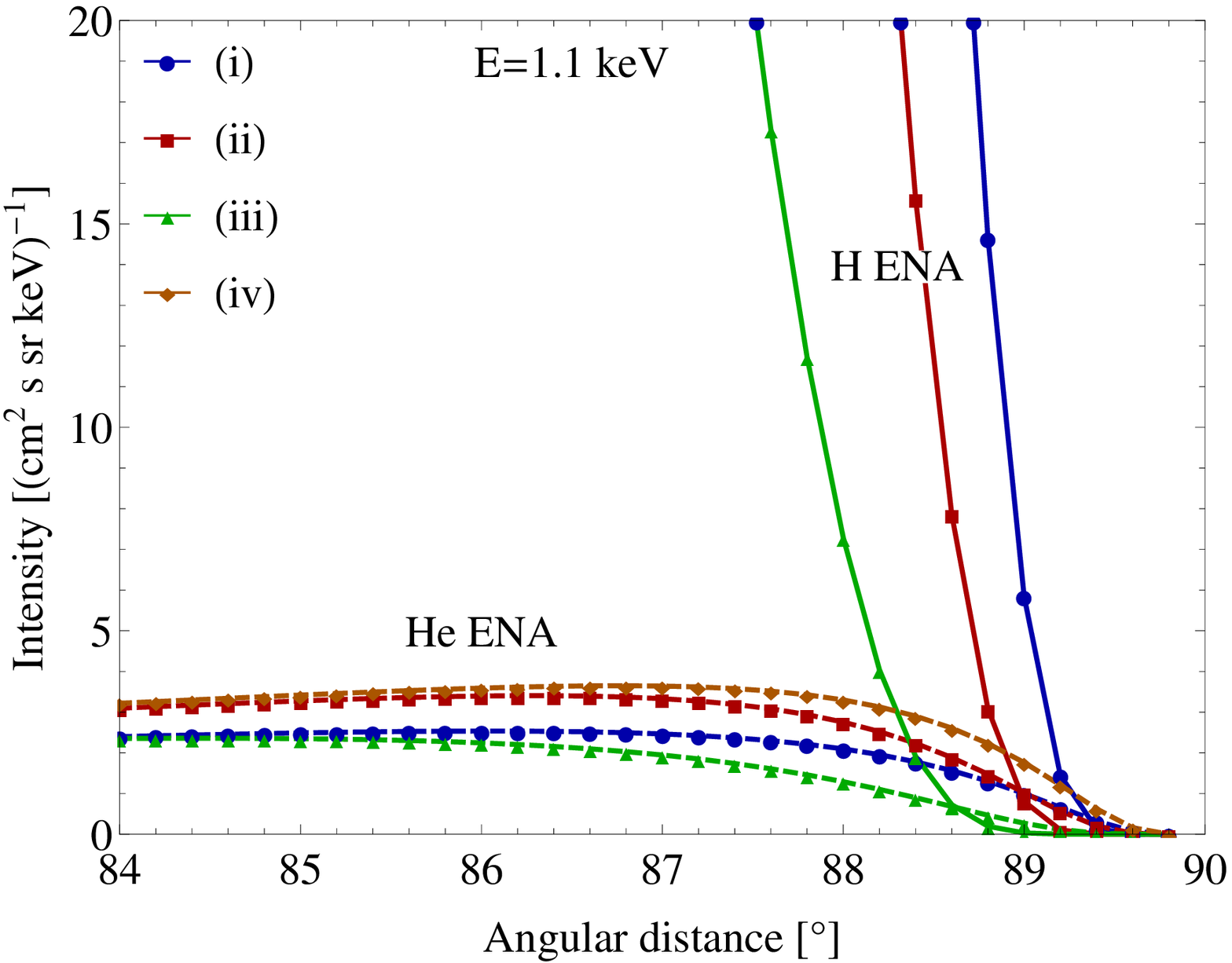}
  \caption{Simulated ENA intensity profiles $I(\theta)$ for $E=1.1~\mathrm{keV}$ in hypothesis (a) (cf.~Section~\ref{sec:grzedz:suprathermal}), solid line for H ENA and dashed for He ENA. Results for parameters sets (i), (ii), (iii), (iv) from Table~\ref{tab:param} are denoted by blue, red, green, and orange (black, darker gray, gray, lighter gray) color, respectively. H ENA intensities for set (iii) and (iv) are the same. In the left panel, He ENA intensities are magnified $\times10$. The right panel is an expanded view of the left panel for high angles.}
  \label{fig:grzedzA}
\end{figure}

\begin{figure}[ht]
  \plottwo{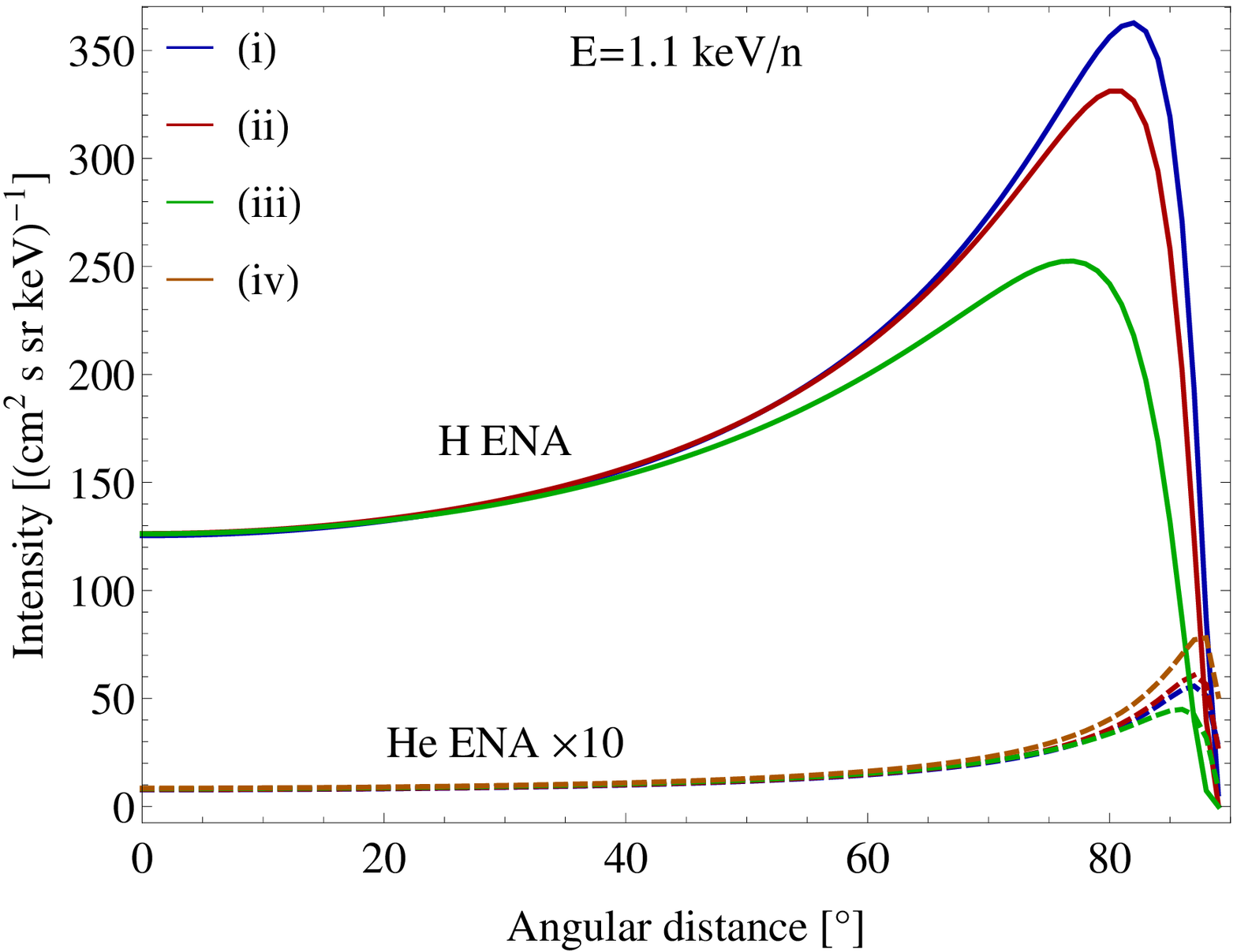}{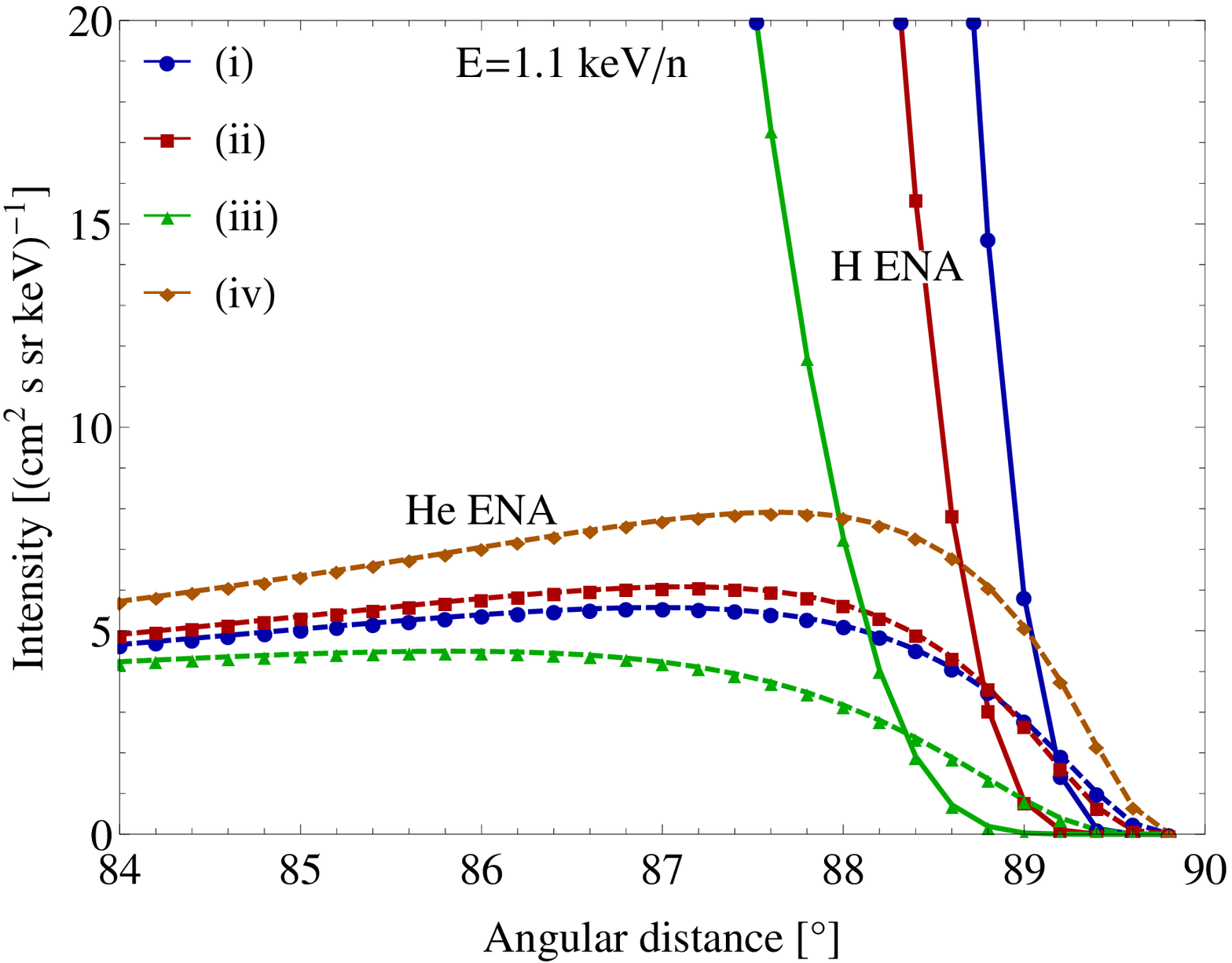}
  \caption{As in Figure~\ref{fig:grzedzA}, but for $E=1.1~\mathrm{keV/n}$ in hypothesis (b) (cf.~Section~\ref{sec:grzedz:suprathermal}).}
  \label{fig:grzedzB}
\end{figure}

The ENA intensity profiles calculated from Equation \eqref{eq:intensity} as a function of angle between the line-of-sight and the direction perpendicular to the interface are presented in Figure~\ref{fig:grzedzA} for hypothesis (a) and in Figure~\ref{fig:grzedzB} for hypothesis (b) (cf.~Section~\ref{sec:grzedz:suprathermal}). The same parameter sets of the LIC ion abundance and the distance to the interface as presented previously by \citet{grzedzielski_etal:10a} are used. They are listed in Table~\ref{tab:param}.

Regardless of the simulation parameters in the extraheliospheric model of the IBEX Ribbon origin, the estimated He ENA intensities seem to be of the order of $\sim(2-7)~\mathrm{(cm^2~s~sr~keV)}$. The highest intensity is expected for the parameters set (iv) in both hypotheses. For these parameters, the ratio of the total extinction thicknesses for helium and hydrogen is the lowest. For other parameters, the relative differences are smaller and depend on different parameter combinations \citep[see Equation 2 in][]{grzedzielski_etal:10a}. Integration of Equation~\eqref{eq:intensity} to get a compact formula is not possible for helium due to the additional factor $q_\mathrm{He^+\to He}(y)$ \eqref{eq:prob}.

Near the edge of the Ribbon, the He ENA signal should be stronger than the signal from H ENA. That is due to smaller extinction thickness for He atoms than for H, which is the main reason for attenuation of the ENA signal near $\theta=90^\circ$. The ratio of the extinction in the LIC for He and H is $\exp\left(\sec\theta(\tau_\mathrm{0,H}-\tau_\mathrm{0,He})\right)$. Thus the ratio of these signals are higher for larger $\theta$ as long as $\tau_\mathrm{0,H}>\tau_\mathrm{0,He}$.

\section{Discussion}

In our previous paper \citep{grzedzielski_etal:13a} we found that the He ENA intensities from the heliosheath in the direction of the IBEX Ribbon in energy $E=(1-5)~\mathrm{keV}$ is $\sim(0.02-0.5)~\mathrm{(cm^2~s~sr~keV)^{-1}}$, depending on the choice of the heliosphere model. For the heliospheric tail direction the intensities are relatively large $\sim(0.5-10)~\mathrm{(cm^2~s~sr~keV)^{-1}}$. Following the model proposed by  \citet{heerikhuisen_etal:10a}, \citet{mobius_etal:13a} derived an analytic model of the IBEX Ribbon. Applying this model to the case of He ENA we obtain that intensity in the same mechanism is very low: $\sim0.014~\mathrm{(cm^2~s~sr~keV)^{-1}}$ (Sec.~\ref{sec:moebius}). This suggests that even if He ENA observation could be done with sufficient accuracy, the He IBEX Ribbon would not be visible. The main reason is the lack of sufficient seed population of neutral solar wind He ENA. This is true for all models where the Ribbon is directly related to the heliosphere.

In the calculation of the He PUI production in the LISM, He ENA produced in the heliosheath were neglected. Also, we assumed that the angle between the line-of-sight and the local magnetic field direction does not change with the distance from the heliosphere. In the case of hydrogen, ENA are produced in a rather thin layer near the heliopause $\lambda_\mathrm{H,LISM}\approx600~\mathrm{AU}$. In the case of helium, that distance is one order of magnitude larger: $\lambda_\mathrm{He,LISM}\approx8000~\mathrm{AU}$. These details do not change the main conclusion that we do not expect Ribbon to be visible in He ENA if Ribbon is created in the \citet{heerikhuisen_etal:10a} mechanism.

Using the extraheliospheric model by \citet{grzedzielski_etal:10a} we found that the He ENA intensity from the IBEX Ribbon direction could be relatively large $\sim(2-7)~\mathrm{(cm^2~s~sr~keV)^{-1}}$ (Sec.~\ref{sec:grzedz}). It seems that such a signal could be distinguishable from the background of He ENA coming from the heliosheath, as estimated in \citet{grzedzielski_etal:13a}. 

It is not straight-forward to replace H by He in the curved version of the extraheliospheric model, because the mean free paths of He ENA in the LIC and the LB are comparable with the curvature radius. In the case of H ENA, the distance to the LB on the line-of-sight tangent to the edge of the spherical interface is much longer than the mean free path in the LIC, so the produced H ENA are almost extinct. Thus the actual curvature near the tangential lines-of-sight is not relevant in this case. For He ENA, a proper treatment of the interface shape should be introduced. Future measurements of the He ENA could give a possibility to determine the curvature of the interface. The spherical model is too crude in the case of He ENA. 

Following \citet{allegrini_etal:08a}, \citet{grzedzielski_etal:13a} suggested that the signal of He ENA could be measured from the tail direction of the heliosphere by the IBEX-Hi detector. Comparable intensity in the direction of IBEX Ribbon would not be measurable because of low He-to-H ENA intensity ratio. The angular size of the Ribbon in the He ENA case would be larger by about $\sim5^\circ$, and the He-to-H ENA intensities ratio would be there greater than 1. However, taking into account that the IBEX field of view is $\sim7^\circ$ wide \citep{funsten_etal:09a}, it seems that the He ENA signal might not be spatially separated from H ENA. 

\begin{figure}[ht]
  \plottwo{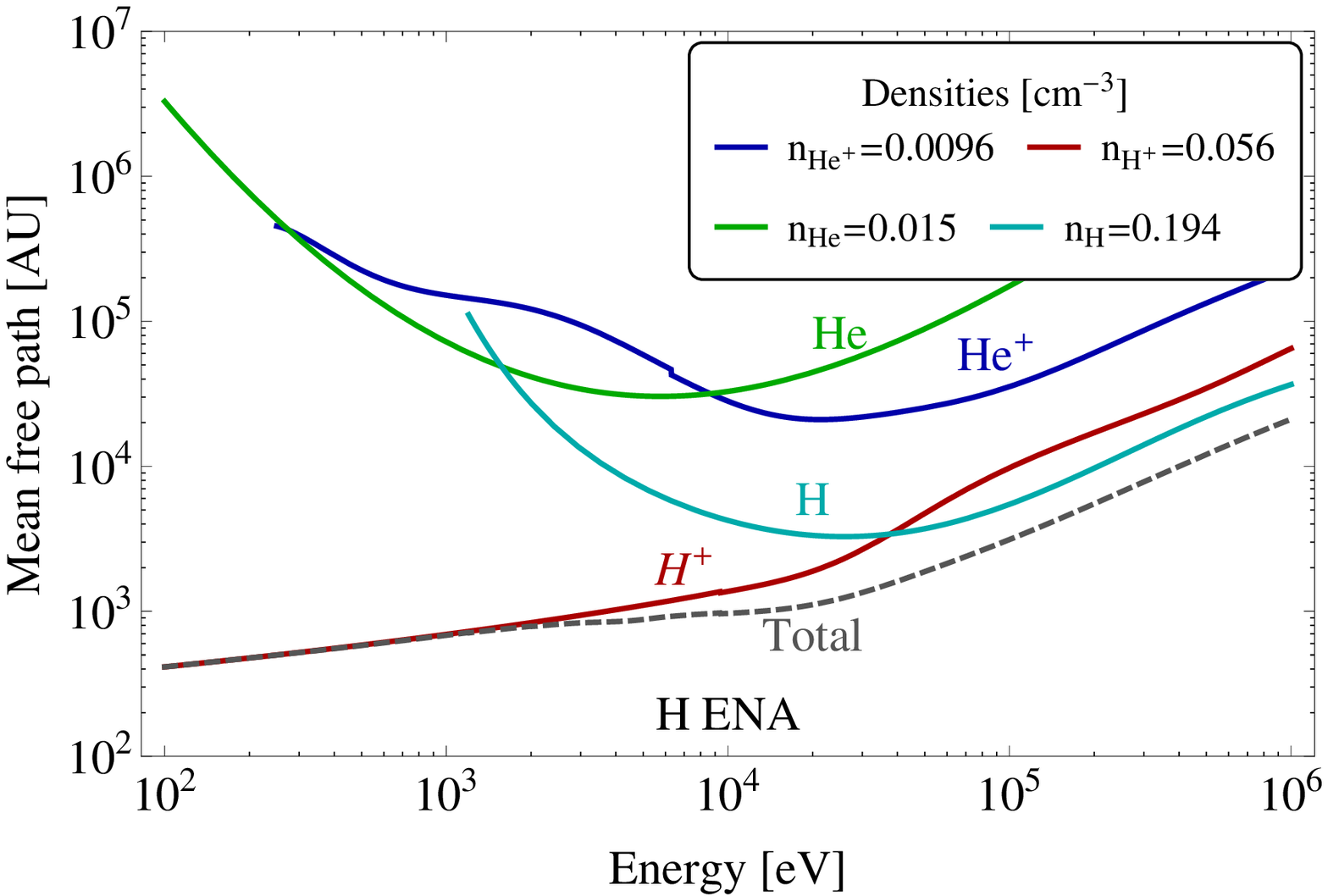}{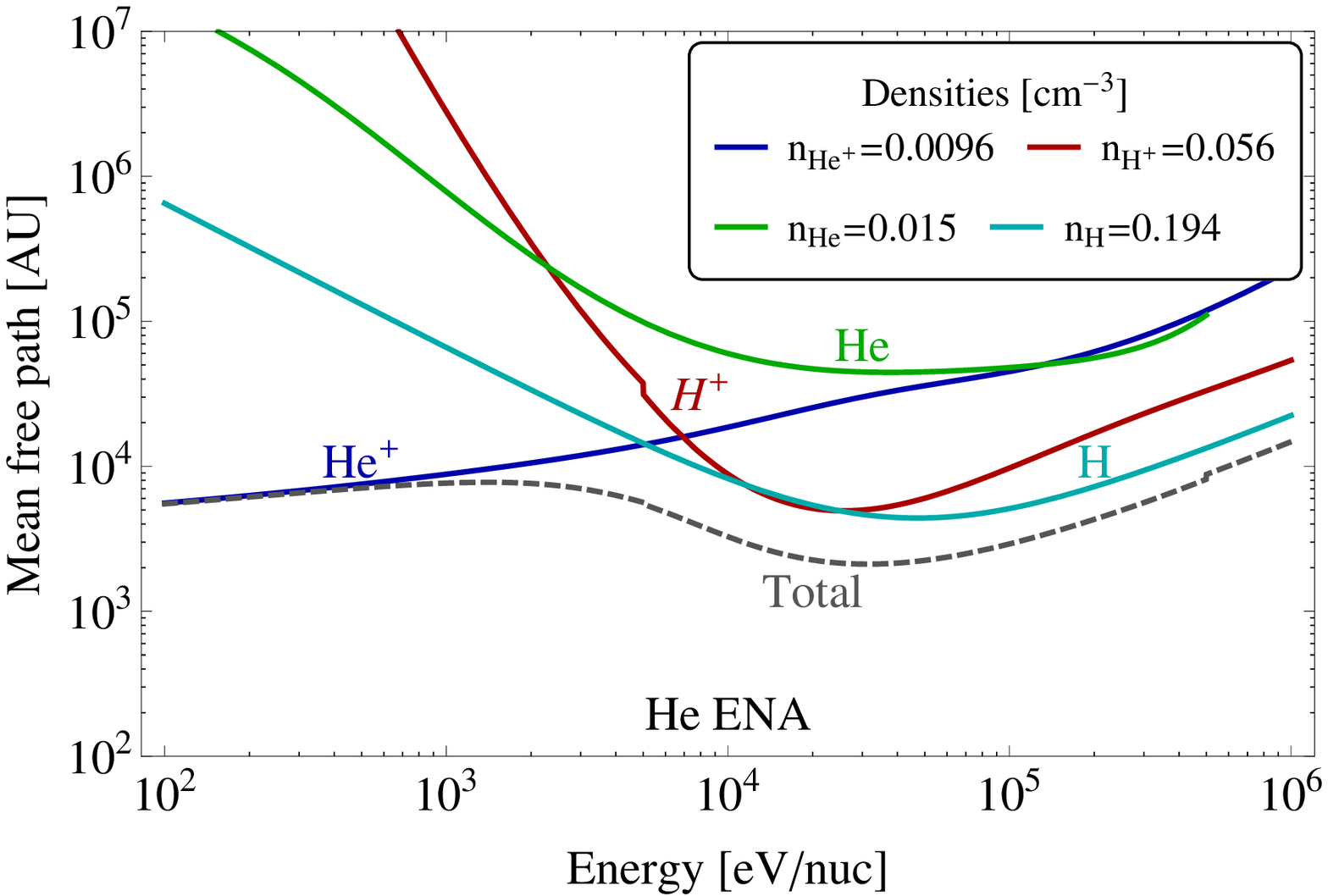}
  \caption{The mean free path for H ENA (left panel) and He ENA (right panel) propagating through the LIC. For H, appropriate cross sections are taken from \citet{lindsay_stebbings:05a} for charge-exchange between H atoms and protons, and from \citet{redbooks} for the others reactions. For He the references are the same as in Table~\ref{tab:grzedz}. The total mean free path (dashed/gray line) is harmonic sum of the contributions (solid lines): $\lambda_\mathrm{mfp}=\left( \sum_i \lambda_i^{-1} \right)^{-1}$. }
  \label{fig:mfp}
\end{figure}

Recent data analysis by \citet{funsten_etal:13a} suggests that the direction to the IBEX Ribbon center depends on the ENA energy channel of the IBEX-Hi detector. Also results obtained by \citet{krimigis_etal:09a} suggest that ENA intensities in higher energies may come from a different source than those for lower energy channels of IBEX. In the extraheliospheric model, the energy dependence on the position of the IBEX Ribbon could be explained by different mean free paths for different energies. Attenuation of intensity at the edge of IBEX Ribbon is caused by the distance to the interface in such a direction. For hydrogen, the mean free path grows with energy monotonically. This is presented in Figure~\ref{fig:mfp}. 
Now we use the densities in the LIC as in Section~\ref{sec:moebius}. Following that it could be possible that the IBEX-Hi signal in the highest energy channel could be produced by a distant source. Also the mean free path for He ENA is large. For energy $E\approx1~\mathrm{keV/n}$, the mean free path is the longest, about $8000~\mathrm{AU}$. It ensures that a He ENA signal from a distant source could be visible even if all H ENA are extinct.

In the currently developed model of the extraheliospheric interface we assume that ENA originated in the LB are partially ionized in the LIC. They become a source for the PUIs in the LIC and could be source of next order ENA due to their neutralization. A contribution from such ``multiple interaction'' ENA should be also added in our model. However, incorporation of such a mechanism in the model requires an appropriate treatment of elastic scattering between atoms and ions, which are now neglected. This mechanism will increase the expected He ENA flux. The results obtained in the present version are still applicable as a lower limit of the expected signal in the extraheliospheric model.

Velocities of He ENA in the analyzed energy range are $v_\mathrm{ENA}\approx (220 - 440)~\mathrm{km~s^{-1}}$, i.e., much higher than thermal velocities of the ions in the LIC $v_\mathrm{H(He),th,LIC}\approx 12.5(6)~\mathrm{km~s^{-1}}$. That justifies using the approximation of relative velocity in the LIC reactions $v_\mathrm{rel}\approx v_\mathrm{ENA}$. However, in the LB the thermal velocity of ions is comparable $v_\mathrm{H(He),th,LB}\approx 160(80)~\mathrm{km~s^{-1}}$, thus we convolved the ENA velocity with the thermal velocity from the Maxwell distributions to get relative velocity $v_\mathrm{rel}$.

\section{Conclusions}

We compare possible He ENA production in two models of the IBEX Ribbon. The first one, developed by \citet{heerikhuisen_etal:10a}, yields the Ribbon He ENA production near the heliopause in the direction where the interstellar magnetic field is perpendicular to the line-of-sight. The second mechanism, proposed by \citet{grzedzielski_etal:10a}, gives production of Ribbon He ENA at the hypothetic nearby contact layer between the LIC and LB. We summarize our main results in the following points.
\begin{itemize}
 \item The He ENA intensity in the direction of IBEX Ribbon was estimated to be of the order of $\sim0.01~ \mathrm{(cm^2~s~sr~keV)^{-1}}$ from the secondary ENA production near the heliopause and is of the same order as the production of He ENA in the heliosheath. Thus, detection of Ribbon in He ENA is not likely in this mechanism.
 
 \item The He ENA intensity in the extraheliospheric model is of order a few $\mathrm{(cm^2~s~sr~keV)^{-1}}$, i.e., it should be easily distinguishable from the He ENA intensity produced in the heliosheath. The angular size of the IBEX He ENA Ribbon is wider than the size in the H ENA case, and near the edge the intensities of He ENA could be even larger than the H ENA intensities. Thus the He ENA Ribbon created by this mechanism should be easily detectable, when instrumental capabilities for He ENA detection of similar quality as the present day IBEX H ENA capabilities become available. 
 
 \item Both models provide He ENA intensities that seem to be below the detectability threshold of the IBEX-Hi detector. Nevertheless, the difference in the expected intensities suggest that a dedicated He ENA instrument would offer a possibility of distinction between the discussed models. 
 
 \item The long mean free path of He ENA in the LIC gives possibility of detection of an interface between the LIC and LB even at distances in excess of a dozen thousand of AU. If the interface is at a larger distance from the Sun, the angular dependence will be qualitatively different because of a higher extinction. Still in the direction perpendicular to the interface ($\theta=0^\circ$) the obtained signal $\sim (0.7-1)~\mathrm{(cm^2~s~sr~keV)^{-1}}$ will be suppressed by the factor of $\exp(-z_0/\lambda_\mathrm{He,LISM})$. Thus the signal should be still over the expected signal from the heliosheath.
 
\end{itemize}

To conclude, the He ENA detection capability would be a welcome improvement of the heliospheric instrumentation because it would offer an interesting insight into the heliospheric physics as well as potential discovery tool for processes operating within $\sim0.1~\mathrm{pc}$ around the heliosphere, unattainable by other measurements methods.  

\acknowledgments
PS \& MB acknowledge support of the Polish National Science Center grant 2012-06-M-ST9-00455.

\bibliographystyle{apj}
\bibliography{ms}

\end{document}